\documentstyle[12pt,aasms4,graphics]{article}

\begin{document}
\title{Severe New Limits on the Host Galaxies of Gamma Ray Bursts}
\author{Bradley E. Schaefer\altaffiltext{1}{schaefer@grb2.physics.yale.edu}
}
\affil{Yale University, Physics Department, JWG 463, New Haven CT
06520-8121, USA}

\begin{abstract}

 The nature of Gamma Ray Bursts (GRBs) remains a complete mystery, 
despite the recent breakthrough discovery of low energy counterparts, although 
it is now generally believed that at least most GRBs are at cosmological 
distances.  Virtually all proposed cosmological models require bursters to
reside in ordinary galaxies.  This can be 
tested by looking inside the smallest GRB error boxes to see if ordinary 
galaxies appear at the expected brightness levels.  This letter reports on
an analysis of the contents of 26 of the smallest regions, many from the
brightest bursts. These events will have $z < 0.4$ and small
uncertainties about luminosity functions, K corrections and galaxy
evolutions; whereas the recent events with optical transients are much 
fainter and hence have high redshifts and grave difficulties in
interpretation. This analysis strongly rejects the many models with peak
luminosities of $10^{57} photons \cdot s^{-1}$ as deduced from the
$LogN-LogP$ curve with no evolution. Indeed, the lower limit on acceptable
luminosities is $6 \times 10^{58} photons \cdot s^{-1}$. The only possible
solution is to either place GRBs at unexpectedly large distances (with $z
> 5.9$ for the faint BATSE bursts) or to require bursters to be far
outside any normal host galaxy.

\end{abstract}


\clearpage
\section{Introduction}

 The discovery of low energy counterparts for Gamma Ray Bursts 
(Costa et al. 1997; van Paradijs et al. 1997; Frail et al. 1997; 
Metzger et al. 1997) has not yet solved the problem of the location of
the burster sites. Measured redshifts associated with optical transients
have values $0.83 < z < 2.1$ (Metzger et al. 1997), $z =  3.42$ (Kulkarni
et al. 1998), and $z =  0.0085$ (Galama et al. 1998), $z = 0.967$
(Djorgovski et al. 1998) for bursts of faint peak flux. Only in the first
and last cases
are the connections between the spectrum and the burst firmly resolved.
The early models of cosmological bursts placed them at distances
corresponding to luminosities of roughly $10^{57} photons \cdot s^{-1}$
(e.g. Fenimore et al. 1993), while it was later realized that the
luminosity could be as high as $2 \times 10^{58} photons \cdot s^{-1}$ if
evolutionary changes are allowed (Horv\'{a}th, M\'{e}sz\'{a}ros, \&
M\'{e}sz\'{a}ros 1996, Totani 1997, Wijers et al. 1998). The candidate
host galaxy for GRB 971214 has a very high redshift $z =  3.42$
(Kulkarni et al. 1998) which suggests that the luminosity could be even as
high as $3 \times 10^{59} photons \cdot s^{-1}$. With nearby, moderate,
and extreme distances all indicated, it is clear that the distance scale
for cosmological bursts is not well established. 

Almost all proposed burst models  place GRBs inside normal galaxies (eg.
Nemiroff 1994). These models can be directly tested by looking 
inside the smallest GRB error regions for the presence of any plausible host 
galaxy.  The recent accurate optical transient positions are much smaller than 
the classical triangulation positions, but they suffer from the faintness of 
the burst and hence the faintness of the expected host, so that the old bright 
bursts are actually more restrictive for the presence of host galaxies.
Indeed, the old bright bursts are at low redshifts where uncertainties in
cosmology and galaxy properties are minimal, while the faint burst with
optical transients are at high redshifts, where K-corrections, cosmology,
luminosity functions, and evolution have large uncertainties.
Nevertheless, the striking result from both old and new GRB positions is the 
stark absence of galaxies at the brightness levels commonly expected.

 This basic no-host-galaxy dilemma was first posed by Schaefer (1992) 
with improvements in analysis by Fenimore et al. (1993) and Woods \& Loeb
(1995).  The problem is that the brightest burst regions (with the smallest
areas) should typically reveal normal galaxies at around sixteenth magnitude 
for the usual distance scales, whereas many of these boxes are empty of 
galaxies to fainter than twentieth magnitude.  Here, a normal galaxy is taken 
as one drawn randomly from the Schechter luminosity function (Binggelli, 
Sandage, \& Tammann 1988) while the usual distance scale (Fenimore et al.
1993; Horv\'{a}th, M\'{e}sz\'{a}ros, \& M\'{e}sz\'{a}ros 1996) has a peak
luminosity of $6 \times 10^{50} erg s^{-1}$ or $10^{57} photon \cdot
s^{-1}$. Band \& Hartmann (1998) introduced a Bayesian analysis procedure
and concluded that an infrared data base (Larson \& McLean 1997) contained
no useful limits while four error boxes observed with the Hubble Space 
Telescope (Schaefer et al. 1997) presented a serious no-host-galaxy problem.

 The previous analysis was based on samples of GRB regions either with 
limits only for the brightest star or galaxy in the field (Schaefer 1992; 
Woods \& Loeb 1995) or with only four regions (Schaefer et al. 1997) or with 
relatively large error boxes for relatively faint bursts (Larson \& McLean
1997).  Schaefer et al. (1998) have accumulated a large data base of 
observations and has placed conservative limits on the U, B, V, R, I, J, H, 
and K magnitudes of the brightest possible galaxies in each of 26 of the 
smallest GRB error boxes. This compilation solves the limitations imposed by 
previous samples and allows for severe new constraints on any host galaxies.  
Table 1 presents these magnitudes on the brightest possible galaxy in each 
field along with some basic properties of the burst (Schaefer et al.
1998). The peak fluxes are for a 0.25 s from 50-300 keV in units of $photons
\cdot s^{-1}$, with almost all provided by E. E. Fenimore in a recent private 
communication.  The magnitudes should be treated as limits since some of the 
values represent detection thresholds and since the host might not be the 
brightest galaxy in a region.  The magnitudes have been corrected 
for extinction in our Galaxy (Zombeck 1990, Blaes et al. 1997).

 Are the observed limits consistent with normal foreground galaxies 
with no causal connection to the GRB?  The incidence of chance galaxies will 
depend only on the area of the error box, the galactic latitude, and the 
filter.  Observed galaxy number densities as a function of magnitude 
(Jones et al. 1991; Smail et al. 1995; Gardner, Cowie, \& Wainscot 
1993) can be used to calculate the magnitude of the brightest expected
foreground galaxy.  In Table 2, the V-band limits on galaxies can be compared 
to the magnitudes for the brightest expected foreground galaxy
($V_{befg}$), with 
a large expected scatter due to the randomness of 
the brightest galaxy and the fact that some of the measures are merely limits 
on the brightest galaxy.  The median of the differences is $0.28 \pm 0.35$
mag for the V-band and $0.07 \pm 0.21 mag$ for all bands.  Thus, the
contents of the 26 GRB error boxes are fully consistent with chance 
foreground galaxies alone.

\section{Analysis}

 A detailed analysis can place limits on the absolute magnitude of any 
host galaxy in each region, for some assumed peak luminosity.  Specifically, 
the observed peak flux can be combined with the assumed peak luminosity to 
yield a luminosity distance to the burster and to its host galaxy.  This 
luminosity distance can then be combined with the observed limits 
on the host magnitude to yield a limit on the absolute magnitude for the host. 
 For the ensemble of boxes, the limits on the absolute magnitudes can then be 
compared to that expected for a normal Schechter luminosity function.  The 
assumed peak luminosity can then be varied until agreement is reached.

 Two classes of distance scales have been widely considered.  The first 
can be called the ``no evolution'' scenario, where the distances are those
whose luminosity corresponds to $6 \times 10^{50} ergs \cdot s^{-1}$
(30-2000 keV) or $10^{57} photons \cdot s^{-1}$ (50-300 keV) as derived
from the $LogN-LogP$ curve (Fenimore et al. 1993; Horv\'{a}th,
 M\'{e}sz\'{a}ros, \& M\'{e}sz\'{a}ros 1996), energetics limits for
compact 
objects, and time dilation (Deng \& Schaefer 1998).  Almost all published 
papers with specific cosmological burst models require and assume this 
distance scale (e.g. Woosley 1993; Usov 1992; Ma \& Xie 1996; Lipunov et
al. 1995; Holdom \& Malaney 1994; M\'{e}sz\'{a}ros \& Rees 1993).
Alternatively, an ``evolutionary'' distance scale might have the GRB number 
density following the rate of massive star 
formation (Totani 1997; Bagot, Zwart, \& Yungelson 1998, Wijers et al.
1998), with
luminosities roughly twenty times larger.  This distance scale is
supported by the recent possible association of a faint GRB with a z= 3.42 
source (Kulkarni et al. 1998).

 The luminosity distance is $D= (L/4\pi P)^{0.5}$, with L the peak
luminosity and P the peak flux.  The limit on the host's absolute magnitude 
is $M =  m - 5 \cdot Log(D)+5$, where $D$ is expressed in parsecs and m is
the limit on the apparent magnitude for any host.  The standard
luminosity-weighted Schechter luminosity function (Binggelli, Sandage, \&
Tammann1988) is adopted with $a =  -1, M^{*} =  -21.0$ (in the V-band for a
Hubble constant of $65 km \cdot s^{-1} \cdot Mpc^{-1}$), and a low
luminosity cutoff at $M =  -14$.
  
 The parameter F is the fraction of the galaxy luminosity function 
which is fainter than the observed limit.  The most critical F measures are 
for bright bursts with small boxes, as these should have bright hosts and few 
foreground interlopers.  We have data from up to seven bands for individual 
bursts, with some being more restrictive while others are less restrictive.
Since all these restrictions simultaneously apply, we can select the minimum 
value, $F_{min}$, as providing the overall limit on the position of the
galaxy within the Schechter luminosity function.  This selection avoids any
penalty associated with including a limit of poor sensitivity in some band.

 We can quantitatively allow for the varying importance of 
large-versus-small boxes and faint-versus-bright bursts by forming a weighted 
average of the individual $F_{min}$ values.  The weight, W, will be the
probability that the brightest galaxy in the field is the host and 
not some foreground galaxy.  This probability is calculated from the magnitude 
of the brightest expected foreground galaxy and its position in the luminosity 
function for the assumed burst luminosity.  The W value does not depend in any 
way on observations of the contents of the region.  The uncertainty in
$<F_{min}>$ will be $[(<F_{min}^{2}>-<F_{min}>^{2})/ \sum W]^{0.5}$.
  
 The weighted average $<F_{min}>$ will (for an assumed peak
luminosity) be a measured statistic for comparison with models.  The model 
$<F_{min}>$ statistic will depend on the existence of hosts.  If normal hosts 
are the brightest galaxy in each of the fields, then the $F_{min}$ values
will be uniformly distributed from zero to one, such that the average of
all 26 values should be 0.5.  With random foreground galaxies, the
observed limits on the individual $F_{min}$ values will be larger, so that
$<F_{min}>$ can only be greater than a half. Similarly, for regions where
a detection threshold is reported, the individual $F_{min}$ values 
can only increase.  Thus, the existence of normal host galaxies requires 
$<F_{min}>$ to be greater than or equal to 0.5.  If hosts are not present in 
the error boxes, then the model $<F_{min}>$ value can vary from near zero
for low L (such that GRBs are relatively nearby and the lack of hosts is
apparent) to near unity for high L (such that GRBs are very distant and the 
lack of hosts is not apparent against the foreground galaxies).  So any 
acceptable model of cosmological GRBs in hosts must adopt a luminosity such 
that $<F_{min}>$ is $\geq 0.5$.

 The analysis must incorporate the effects of the red shift on the 
observed brightness of the burst and of the host galaxy.  K-corrections plus 
E-corrections for a distribution of galaxy types have been adopted from 
evolutionary synthetic spectral models (Rocca-Volmerange \& Guideroni
1988; Pozetti, Bruzual, \& Zamorani 1996).  K-corrections for 
the bursts have been calculated following equations 1, 2, and 4 in 
Fenimore et al. (1992).  I have adopted an average spectral slope index of 
-1.5 (see Figure 46 of Schaefer et al. 1994), although this value is varied 
as described below.  The use of a power law spectral model is acceptable, 
since the $<F_{min}>$ value is insensitive to large changes in the slope
(cf. Fenimore \& Bloom 1995). This procedure corrects the observed peak
fluxes from 50 to 300 KeV for the effect of redshift. The Hubble Constant
and the deceleration
parameter enter the problem for the value of $M^{*}$ as well as to
calculate the E-corrections from the luminosity distance.  I have adopted 
$H_{0} =  65 km \cdot s^{-1} \cdot Mpc^{-1}$ and $q_{0}= 0.5$, although
these values have also been varied.

\section{No Evolution Case}

 With these generalizations, we can first address the ``no
evolution'' peak luminosity of $10^{57} photon \cdot s^{-1}$ ($\sim 6
\times 10^{50} ergs \cdot s^{-1}$). Table 2
presents the derived red shift (z), the most restrictive band 
(MRB), $F_{min}$, and W for all 26 bursts.  The weighted average
$<F_{min}>$ is $0.141 \pm 0.053$.
  
 How robust is this result?  (1) The $<F_{min}>$ value varies from
0.153 to 0.141 as the deceleration parameter varies from 0.1 to 0.5.  
As the assumed Hubble Constant is varied from 50 to 80 $km \cdot s^{-1}
\cdot Mpc^{-1}$, the $<F_{min}>$ value changes from 0.125 to 0.180.  (2)
How large a change in the luminosity function parameters is needed to satisfy 
the limits?  The $<F_{min}>$ value will increase to a half by either
making the low luminosity slope equal -2.1 or the $M^{*}$ value over 5
magnitudes fainter.  The low luminosity cutoff is unimportant for a 
luminosity-weighted function.  (3) If only the V-band data are considered,
the $<F_{min}>$ value is $0.231 \pm 0.063$.  (4) If there were some
[totally unsuspected] systematic error that brightened the limits in Table 1, 
any such errors would have to average 2.5 magnitudes to get $<F_{min}>$
greater than 0.5.  (5) If the three 1997 bursts are arbitrarily ignored
, then the $<F_{min}>$ value is $0.133 \pm 0.051$. The inclusion of
GRB971214 into Table 1 (with $R > 25.6$, $P_{256} \sim 2 ph \cdot
s^{-1}$ and $z =  3.42$) changes the $<F_{min}>$ slightly to $0.155
\pm 0.055$. (6) If host galaxies only occupy some fraction of the error 
boxes, then $<F_{min}>$ will vary linearly between the host+foreground
level of 0.55 and the 
foreground alone level of 0.19.  For an observed $<F_{min}>$ of
$0.141 \pm 0.053$, the two-sigma acceptable value can be modeled by requiring 
that $>84 \%$ of the boxes do not have hosts.  (7) If no K-corrections for
the host galaxy are used, then $<F_{min}> =  0.136 \pm 0.054$.  As the
average GRB spectral slope index changes from 1.0 to 2.5, the $<F_{min}>$
value varies over a range of amplitude 0.022. (8) The analysis never uses
a burst rate density so the result is independent of any assumptions on
the rate density evolution.

 What about the possibility of a luminosity function for the bursts?  
The effect on the $<F_{min}>$ statistic will be to average it over the
assumed luminosity function.  To get an expected $<F_{min}>$ greater than
a half, the majority of the bursts must greatly exceed the luminosity.  So
a GRB luminosity function cannot solve the basic no-host-galaxy dilemma.

 Let me state the no-host-galaxy dilemma in five ways with increasing 
generality: (1) The smallest classical GRB box is for GRB790406 with
$P =  45 photon s^{-1} \cdot cm^{-2}$, so that $z =  0.09$ for the ``no
evolution'' luminosity and an $M^{*}$ galaxy should appear as $B =  17.8$
mag.  Yet the region is empty of galaxies to $B =  24.29$ mag, so that any
host must be $>6.5$ mag fainter than $M^{*}$.  (2) The existing limits on
the hosts for GRB970228 and GRB970508 require the hosts to be in the bottom
 $0.3\%$ and the bottom $2.2\%$ (see $F_{min}$ values in Table 2).  For
such faint galaxies, the luminosity 
function is not well known, yet it is well known enough to realize that both 
hosts are improbably faint were they to be normal galaxies.  (3) For the larger
 sample of GRBs with $W > 0.9$, the average $F_{min}$ values are very low,
with nine of the fourteen events whose hosts must be in the bottom $4\%$ 
of the luminosity function.  (4) The brightest galaxy in the 26 regions has a 
median difference from the brightest expected foreground galaxy brightness by 
$0.07 \pm 0.21$ mag, showing that the contents of the error boxes are
entirely consistent with random foreground galaxies.  (5) The $<F_{min}>$
value is $0.141 \pm 0.053$ and there is no plausible means to make it
$\geq 0.5$.

\section{Evolutionary Case}

 What about the possibility that the peak luminosity is substantially 
brighter than the ``no-evolution'' value?  Just such a case is expected if
the GRB number density follows the rate of massive star formation 
(Totani 1997).  The combined BATSE and PVO $LogN-LogP$ curve can be made
consistent with average L values up to $2 \times 10^{58} photon \cdot
s^{-1}$ ($\sim 10^{52} ergs \cdot s^{-1}$) for a careful choice of density
evolution and luminosity function 
(Horv\'{a}th, M\'{e}sz\'{a}ros, \& M\'{e}sz\'{a}ros 1996).  An equivalent
way to quantify this limit is with the red shift of the BATSE $90 \%$ 
efficiency threshold ($z_{0.85}$ for $P_{256} =  0.85 photon \cdot
s^{-1}$), with values ranging from two to three (Totani 1997).  This 
luminosity from the evolutionary scenario can be tested against the limits on 
host galaxies.  

 Table 2 presents the values of $z$, $MRB$, $F_{min}$, and W for
all 26 bursts on the assumption that $L =  2 \times 10^{58} photon \cdot 
s^{-1}$ (with $z_{0.85}= 3.2$).  The $<F_{min}>$ value is $0.291 \pm 0.118$ 
(versus 0.55 expected for normal host+foreground), with most of the
 information coming from five bursts with small boxes.  All but one of the
 bursts have z around a half, so that K- corrections and luminosity
functions are still known with some confidence.  The $<F_{min}>$ is 
inconsistent with the presence of host galaxies in the GRB regions at the
2.2-sigma confidence level.  For $<F_{min}>$ to be greater then a half, L
must be greater than $10^{59} photon s^{-1}$ (with $z_{0.85}= 7.9$),
although a luminosity of $6 \times 10^{58} photon \cdot s^{-1}$ (with
$z_{0.85} =  5.9$) is at the one-sigma limit. The addition of GRB971214
only slightly changes $<F_{min}>$ to $0.323 \pm 0.106$. This represents a
conservative 
limit since (1) the brightest galaxy in the box might not be the host, (2)
half the relevant magnitude limits merely represent detection thresholds,
and (3) the host+foreground case predicts $<F_{min}>$ equal to 0.55.

Uncertainties rise as the hosts are pushed to farther distances. For
example, the effects of uncertainties in the cosmological parameters
increase, the role of dust in obscuring young galaxies could perhaps
become important, and the luminosity function might change significantly. 
Fortunately, the bursts used in this study are very bright (the median for
bursts with $W > 0.1$ is $P_{256} = 45 photons \cdot cm^{-1} \cdot
s^{-1}$) and hence close, and so cosmological uncertainties are small,
there is no abnormal dust obscuration, and the luminosity function is
substantially unchanged (Ellis et al. 1996). In constrast, the bursts with
transients are systematically fainter (the median for the nine bursts is
$P_{256} = 3.3 photons \cdot cm^{-1} \cdot s^{-1}$) and hence farther away
by a factor of $\sim 3$, and so have many more problems caused by
cosmology, dust and evolution. This crucial difference is why host galaxy
limits from the bright bursts are more constricting than limits from the
faint bursts with small boxes. 

\section{Possible Solutions}

 If GRBs are cosmological, then there must be some solution to the 
no-host-galaxy dilemma.  I can only think of two classes of solutions, first 
where the bursts are placed at very large cosmological distances and second
where bursters do not reside in normal hosts galaxies.

 If GRBs are placed at extreme distances, then the required peak 
luminosity is $L > 6 \times 10^{58} photon \cdot s^{-1}$ with the BATSE
faint bursts at $z_{0.85} > 5.9$.  Any such model would have 
to fine tune the cosmology and density evolution to produce the long -3/2
slope region of the PVO $LogN-LogP$ curve (Fenimore et al. 1993).  Any
such model would have trouble explaining the $z < 2.1$ limit for GRB970508
(Metzger et al. 1997) and the $z = 0.967$ redshift for
GRB980703 (Djorgovski et al. 1998). Any such model is inconsistent (Deng
\& Schaefer 1998) with the observed time  dilation of burst light curves
(Norris et al. 1994, in't Zand \& Fenimore 1996, Deng \& Schaefer 1998).
Finally, any such model places bursts at distances already rejected by
limits on gravitational lensing (Marani 1998).

 GRBs might not reside in normal galaxies for various reasons.  It 
might be that bursters were ejected from their galaxy of origin at high 
velocity so as to now appear far away.  But any ejection mechanism must be
$>84\%$ efficient .  Also, an analysis of six high-latitude bright bursts with 
small boxes shows that the area around the box is empty, forcing 
the average ejection-to-burst time to be $>2 \times 10^{9}$ years for
ejection velocities of $500 km \cdot s^{-1}$ for the canonical peak
luminosity.  A second alternative is that GRBs occur with equal 
probability per galaxy regardless of mass.  Yet such a possibility is 
formally rejected (with $<F_{min}>= 0.45$) while even models involving
compact objects in galactic centers still have 
burst frequency being mass dependent.  A third possibility is that the hosts 
are of a greatly subluminous population, with a luminosity function that has 
$M^{*}$ fainter by 4.9 mag.  Such an ad hoc assumption would require 
identifying an appropriate population and explaining 
why normal galaxies do not produce bursts.  A final alternative, is that GRBs 
are indeed in intergalactic space, yet then there is no known source of 
compact objects of the required energy.

 In conclusion, Gamma Ray Bursters are strongly shown to not reside in 
normal host galaxies at either the ``no evolution'' distance scale
($L =  10^{57} photon \cdot s^{-1}$ and $z_{0.85} =  0.69$) or the
``evolutionary'' distance scale associated with bursts as tracers of star
formation ($L =  2 \times 10^{58} photon \cdot s^{-1}$ and $z_{0.85}= 3.2$).
This no-host-galaxy dilemma rejects many models, and forces GRBs to either 
be at very large distances ($L >  6 \times 10^{58} photon \cdot s^{-1}$ 
with the BATSE faint bursts at $z > 5.9$) or to not be in normal host
galaxies.

\clearpage

\begin {table}
\begin {center}
\begin {tabular}{|cccc|cccccccc|}

\hline

&$b_{II}$ & Area & $P_{256}$ & \multicolumn{8}{|c|}{Apparent Magnitude
Limits (Extinction Corrected)} \\ 
GRB & $^{o}$ & sq' & $ph \cdot cm^{-2} \cdot s^{-1}$ & U & B & V & R & I &
J & H &
K\\

\hline
781104 & -26 &14 &69 &16.51 &15.94 &15.02 &... &14.04
&... &... &... \\
781119 &-84 &8 &63 &19.34 &20.16 &19.68 &18.20 &...
&... &... &... \\
781124 &80 &48 &99 &20.90 &20.62 &18.69 &... &17.81
&... &... &... \\
790113 &-19 &78 &105 &... &18.44 &17.18 &16.03 &14.53
&... &... &... \\
790307 &14 &10 &57 &... &18.91 &18.40 &... &17.93
&... &... &... \\
790313 &-25 &24 &39 &... &17.85 &17.20 &... &16.50
&... &... &... \\
790325 &22 &2 &29 &22.74 &22.11 &20.69 &19.46 &...
&... &... &13.56 \\
790329 &57 &41 &19 &20.37 &19.66 &18.19 &... &16.72
&... &... &... \\
790331 &-6 &20 &48 &... &... &13.70 &13.53 &13.05
&... &... &... \\
790406 &-61 &0.26 &45 &22.46 &24.29 &23.26 &22.22 &21.53
&17.49 &... &18.30 \\
790418 &-16 &2.9 &46 &... &... &21.71 &... &20.00
&18.98 &18.21 &17.53 \\
790613 &38 &0.76 &15 &20.92 &21.06 &20.17 &19.45 &19.33
&... &... &13.58 \\
791105 &-53 &35 &21 &20.16 &20.08 &19.27 &... &19.31
&... &... &... \\
791116 &-75 &3.7 &69 &22.11 &22.55 &21.12 &19.99 &19.10
&... &... &... \\
910122 &-31 &19.3 &42 &... &20.58 &19.33 &... &17.95
&16.86 &... &15.28 \\
910219 &55 &7.29 &32 &... &... &19.81 &19.08 &18.32
&16.99 &16.39 &15.59 \\
911118 &36 &12.2 &41 &20.06 &20.17 &18.89 &18.31 &17.60
&... &... &... \\
920325 &-44 &2.1 &25 &... &20.39 &20.42 &... &...
&... &... &16.99 \\
920406 &-28 &0.44 &73 &22.27 &23.42 &21.99 &21.84 &21.39
&19.35 &... &18.27 \\
920501 &1 &0.89 &48 &... &6.41 &... &14.09 &...
&16.57 &... &15.25 \\
920711 &27 &1.4 &23 &... &20.80 &20.38 &20.33 &...
&19.05 &18.16 &17.47 \\
920720 &81 &1.3 &43 &... &18.54 &18.04 &... &...
&16.40 &15.90 &15.40 \\
920723 &8 &4.46 &95 &... &... &16.88 &... &16.30
&17.82 &... &16.04 \\
970228 &-18 &0.003 &10 &... &... &24.63  &...  &24.19
&... &... &... \\
970402 &-9 &2.2 &0.5 &18.81 &18.11 &17.41 &17.05 &16.79
&... &... &... \\
970508 &26 &0.003 &1.2 &... &26.36 &25.56 &25.37 &24.01
&... &20.25 &... \\
\hline
\end{tabular}
\end{center}
\caption{Limits on Brightest Galaxy}
\end {table}

\begin {table}
\begin {center}
\begin {tabular}{|ccc|cccc|cccc|}

\hline  
&&& \multicolumn{4}{|c|}{$L= 10^{57} photon \cdot s^{-1}$} &
\multicolumn{4}{|c|}{$L= 2\times10^{58} photon \cdot s^{-1}$} \\
\hline  
GRB          & $V$ &$V_{befg}$ &z &MRB &$F_{min}$ &W
&z &MRB &$F_{min}$ &W\\ 
\hline

781104 &15.02 &18.93 &0.08 &U &0.985 &0.86 &0.34 &U
&1.000 &0.00 \\
781119 &19.68 &19.47 &0.08 &V &0.075 &0.91 &0.36 &V
&0.888 &0.07 \\
781124 &18.69 &17.74 &0.06 &B &0.039 &0.77 &0.29 &U
&0.716 &0.00 \\
790113 &17.18 &17.28 &0.06 &B &0.249 &0.68 &0.28 &B
&1.000 &0.00 \\
790307 &18.40 &19.26 &0.08 &I &0.230 &0.63 &0.38 &I
&0.996 &0.00 \\
790313 &17.20 &18.41 &0.10 &V &0.748 &0.19 &0.45 &I
&1.000 &0.00 \\
790325 &20.69 &20.81 &0.12 &U &0.033 &0.93 &0.53 &U
&0.515 &0.19 \\
790329 &18.19 &17.90 &0.14 &U &0.366 &0.25 &0.65 &U
&1.000 &0.00 \\
790331 &13.70 &18.59 &0.09 &V &1.000 &0.25 &0.41 &V
&1.000 &0.00 \\
790406 &23.26 &22.78 &0.09 &B &0.002 &1.00 &0.42 &B
&0.107 &0.82 \\
790418 &21.71 &20.45 &0.09 &V &0.017 &0.95 &0.42 &J
&0.315 &0.16 \\
790613 &20.17 &21.75 &0.17 &B &0.212 &0.94 &0.73 &U
&0.993 &0.07 \\
791105 &19.27 &18.05 &0.14 &I &0.180 &0.01 &0.62 &I
&0.978 &0.00 \\
791116 &21.12 &20.22 &0.08 &B &0.009 &0.96 &0.34 &B
&0.292 &0.28 \\
910122 &19.33 &18.62 &0.10 &B &0.098 &0.77 &0.44 &J
&0.944 &0.00 \\
910219 &19.81 &19.56 &0.11 &V &0.139 &0.83 &0.50 &J
&0.969 &0.00 \\
911118 &18.89 &19.07 &0.10 &B &0.139 &0.83 &0.44 &B
&0.996 &0.00 \\
920325 &20.42 &20.76 &0.13 &V &0.107 &0.93 &0.57 &K
&0.864 &0.00 \\
920406 &21.99 &22.27 &0.07 &B &0.003 &1.00 &0.33 &B
&0.139 &0.83 \\
920501 &... &... &0.09 &J &0.057 &0.97 &0.41 &J
&0.682 &0.52 \\
920711 &20.38 &21.16 &0.13 &J &0.036 &0.90 &0.59 &J
&0.483 &0.14 \\
920720 &18.04 &21.23 &0.10 &J &0.195 &0.95 &0.43 &J
&0.985 &0.35 \\
920723 &16.88 &20.04 &0.06 &J &0.024 &0.94 &0.29 &J
&0.422 &0.28 \\
970228 &24.63 &27.16 &0.20 &I &0.003 &1.00 &0.90 &I
&0.082 &0.99 \\
970402 &17.41 &20.72 &0.90 &U &1.000 &0.00 &4.32 &U
&1.000 &0.00 \\
970508 &25.56 &27.16 &0.58 &R &0.022 &1.00 &2.70 &B
&0.117 &0.98 \\

\hline
\end{tabular}
\end{center}
\caption{$F_{min}$ values.}
\end {table}
\end{document}